\documentclass[useAMS]{mn2e}
\usepackage{graphicx}
\usepackage{epsfig}
\usepackage{amssymb}
\usepackage{lscape}
\usepackage{ulem}
\usepackage{txfonts}

\def\kms{km~s$^{-1}$}

\def\c2s{C\,{\sc ii}$^{\star}$}
\def\hkpc{$h_{70}^{-1}$ kpc}

\def\ebmv{E$(B-V)$}

\def\dv{$\Delta$V}

\def\lsun{L$_{\odot}$}
\def\msun{M$_{\odot}$}
\def\rp{$r_p$}

\def\LIR{L$_{IR}$}

\title[The LIRG-merger connection] {Galaxy pairs in the Sloan digital sky survey - VII:
The merger -- luminous infra-red galaxy connection.}

\author[Ellison et al.] {Sara L. Ellison$^1$, J. Trevor Mendel$^1$,
Jillian M. Scudder$^1$, David R. Patton$^2$, \newauthor Michael
J. D. Palmer$^1$\\ $^1$ Department of Physics \& Astronomy, University
of Victoria, Finnerty Road, Victoria, British Columbia, V8P 1A1,
Canada.\\ 
$^2$ Department of Physics \& Astronomy, Trent University,
1600 West Bank Drive, Peterborough, Ontario, K9J 7B8, Canada.  }

\begin{document}

\maketitle

\begin{abstract}
We use a sample of 9397 low redshift ($z\le0.1$) galaxies
with a close companion to investigate the connection between mergers and
luminous infra-red (IR) galaxies (LIRGs).  The pairs are selected from
the Sloan Digital Sky Survey (SDSS) and have projected separations
\rp\ $\le$ 80 \hkpc, relative velocities \dv\ $\le$ 300 \kms\ and stellar
mass ratios within a factor 1:10.  A control sample consisting
of 4 galaxies per pair galaxy is constructed by simultaneously matching
in stellar mass, redshift and environment to galaxies with no close
companion.  The IR luminosities (\LIR) of galaxies in the pair and
control samples are determined from the SDSS -- Infra-red Astronomical
Satellite (IRAS) matched catalog of Hwang et al. (2010).  Over the
redshift range of our pairs sample, the IRAS matches are complete to
LIRG luminosities (\LIR $\ge 10^{11}$ \lsun), allowing us to
investigate the connection between mergers and luminous IR galaxies.
We find a trend for increasing LIRG fraction towards smaller pair
separations, peaking at a factor of $\sim$ 5--10 above the median control
fraction at the smallest separations ($r_p < 20$ \hkpc), but remaining 
elevated by a
factor $\sim$ 2--3 even out to 80 \hkpc\ (the widest separations in our
sample).  LIRG pairs predominantly
have high star formation rates (SFRs), high extinction and are found
in relatively low density environments, relative to the full pairs
sample.  We also find that LIRGs are most likely to be found in high
mass galaxies which have an approximately equal mass companion.  We
confirm the results of previous studies that both the active galactic
nucleus (AGN) fraction and merger fraction increase strongly as a
function of IR luminosity.  About 7\% of LIRGs are associated with
major mergers, as defined within the criteria and mass completion
of our sample.  Finally, we quantify a SFR offset ($\Delta$SFR) as the
enhancement (or decrement) relative to star-forming galaxies of the
same mass and redshift.  We demonstrate that there is a clear
connection between the $\Delta$SFR and the classification of a galaxy
as a LIRG that is mass dependent.  Most of the LIRGs in our merger
sample are relatively high mass galaxies (log (M$_{\star}$/\msun) 
$> 10.5$), likely because the SFR enhancement required to produce LIRG
luminosities is more modest than at low masses.  The $\Delta$SFR
offers a redshift-independent metric for the identification of the
galaxies with the most enhanced
star forming rates that does not rely on fixed \LIR\ boundaries.

\end{abstract}

\begin{keywords}
Galaxies: interactions, infrared: galaxies, galaxies: active
\end{keywords}

\section{Introduction}

The total far infra-red (IR) luminosity (\LIR), which is defined as
the bolometric contribution from 8-1000 $\mu$m, of a galaxy is closely
tied to its star formation rate (SFR).  The correlation is due to the
processing of UV photons by galactic dust, which are re-emitted in the
thermal spectrum (see, for example, the reviews on galactic star
formation by Kennicutt 1998; Kennicutt \& Evans 2012).  There can be
additional contributions to \LIR\ from an active galactic nucleus
(AGN).  However, most studies find that for the general population,
the AGN contribution to the total \LIR\ is modest, at about 10--20\%,
although the AGN contribution can become dominant at the highest far-IR
luminosities (Nardini et al. 2010; Yuan, Kewley \& Sanders 2010;
Petric et al. 2011; Lee et al. 2012).

Based on the two mechanisms for IR emission, it can be easily
appreciated why mergers provide a fertile environment for high IR
luminosities. It is well documented that mergers have, on average,
enhanced SFRs relative to isolated counterparts (e.g. Kennicutt et
al. 1987; Barton, Geller \& Kenyon 2000; Bergvall et al.  2003; Lambas
et al. 2003; Nikolic, Cullen \& Alexander 2004; Alonso et al. 2004,
2006; Li et al. 2008; Ellison et al. 2008, 2010; Woods et al. 2010;
Darg et al. 2010a; Patton et al. 2011).  Although typical SFR
enhancements are a factor of a few at both low and high redshift
(Robaina et al. 2009; Wong et al.  2011), a minority of galaxies can
have SFR enhancements of a factor of ten or more (Scudder et
al. 2012).  Interactions can also lead to nuclear activity,
triggering AGN over a wide variety of energies from Seyferts to
quasars (Kennicutt et al. 1987; Ramos-Almeida et al. 2011, 2012;
Canalizo \& Stockton 2001; Ellison et al 2011b; Silverman et al. 2011;
Koss et al. 2012).  It is therefore not surprising that pairs also
show increased \LIR\ and that interactions can contribute
significantly to the IR luminosity (e.g.  Lonsdale, Persson \&
Matthews 1984; Kennicutt et al 1987; Telesco, Wolstoncroft \& Done
1988; Xu \& Sulentic 1991; Xu et al. 2000; Hernandez Toledo,
Dultzin-Hacyan, \& Sulentic 2001; Bridge et al. 2007; Smith et
al. 2007; Hwang et al. 2010, 2011).

At the highest IR luminosities, galaxies are classified as luminous IR
galaxies (LIRGs) and ultra-luminous IR galaxies (ULIRGs) when their
total far-IR luminosities are $11 \le$ log (\LIR/\lsun) $< 12$ and log
(\LIR/\lsun) $\ge$12 respectively (see Sanders \& Mirabel 1996
and Veilleux et al. 2002 for reviews).  It is well known that, at
least at low redshift, the vast majority of ULIRGs are mergers, with
recent work suggesting that mergers may also contribute significantly
to ULIRGs at high $z$ (e.g. Dasyra et al. 2008; Kartaltepe et
al. 2010, 2012).  As the \LIR\ decreases, so does the fractional
contribution from mergers (e.g. Kartaltepe et al. 2010), such that at
log (\LIR/\lsun) = 11.5 about 20\% of LIRGs are major mergers.
Moreover, at fixed \LIR, there is an increasingly important
contribution from mergers at lower redshifts (Melbourne, Koo \& Le
Floc'h 2005; Wang et al. 2006).  However, not all mergers lead to such
prodigious IR emission.  Statistically this is obvious from the
comparison of the ULIRG space density of $\sim 10^{-7}$ Mpc$^{-3}$ at
low redshift (Kim \& Sanders 1998), compared to the space density of
mergers (e.g. Chou, Bridge \& Abraham 2011) which is 3 orders of
magnitude higher.

In this paper, we study the IR properties of a large
sample of low redshift close pairs in order to further
investigate the connection between high \LIR\ and mergers.
Whilst the fraction of LIRGs that are mergers has been previously
quantified, the fraction of mergers that result in LIRGs is
not well constrained.  How frequently does a merger result in a LIRG,
and can we identify \textit{which} mergers are most conducive to
LIRG production?  We also compare the properties of the LIRGs in
close galaxy pairs with LIRGs in our control sample, in order to
understand whether merging plays a special role in LIRG production,
or whether interactions are simply one pathway to prodigious IR
emission.

\medskip

We adopt a cosmology of  $\Omega_{\Lambda} = 0.7$, $\Omega_M = 0.3$,
$H_0 = 70$ km/s/Mpc.

\section{Sample selection}\label{sample_sec}

\subsection{Infra-red selection}

The study presented here combines a spectroscopically selected sample
of pairs from the Sloan Digital Sky Survey (SDSS) with IR luminosities
obtained from the Infra-red Astronomical Satellite (IRAS).  IRAS
surveyed the sky in four bands at 12, 25, 60 and 100 $\mu$m.  Hwang et
al. (2010, hereafter H10) matched the IRAS Faint Source Catalogue
(FSC), which contains 173, 044 IR sources at $|b|>10$ deg with the
SDSS data release (DR) 7.  H10 require that IRAS sources must have
robust detections at 60 $\mu$m, but no requirements are placed on
detections in the other bands. Bolometric IR luminosities are
calculated by H10 via spectral energy distribution (SED) fits to the
templates of Chary \& Elbaz (2001) at 60 and (where available) 100
$\mu$m.  Hwang et al. (2010) show that IRAS 60 $\mu$m only
determinations of \LIR\ agree well with those derived from 60+100
$\mu$m SED fits, with a scatter of only 0.08 dex (see also Zhu et
al. 2008).

\begin{figure}
\centerline{\rotatebox{270}{\resizebox{6cm}{!}
{\includegraphics{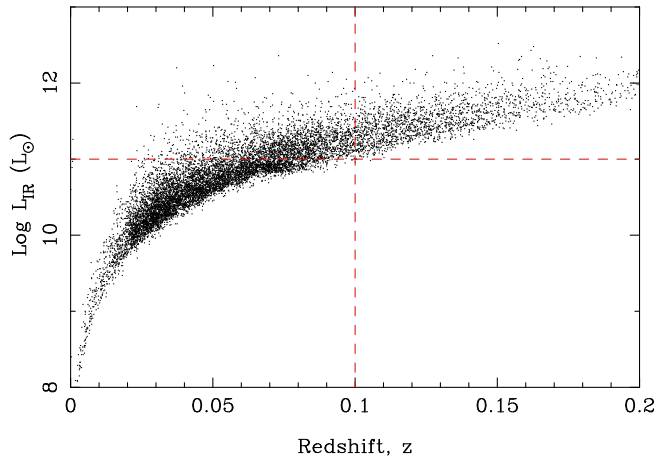}}}}
\caption{\label{z_lir}\LIR\ versus redshift for IRAS-detected galaxies 
matched to the SDSS DR7 in the Hwang et al. (2010) catalog. 
We adopt a redshift cut $z\le$0.1 (vertical dashed line)
so that we are complete for LIRG 
luminosities of log (\LIR/\lsun) $\ge$ 11 (horizontal dashed line).}
\end{figure}

One of the main goals of this work is to study the properties and
frequency of LIRGs in our galaxy pair sample.  Figure \ref{z_lir}
shows the distribution of redshift and \LIR\ for galaxies in the H10
catalog, and demonstrates the increasing incompleteness at small
\LIR\ at larger distances.  In order to be complete down to the
limiting LIRG luminosity (log (\LIR/\lsun) $\ge$ 11), we impose a
redshift cut $z\le0.1$.

\subsection{Close pairs sample}\label{pairs_sec}

We select a sample of close galaxy pairs from the subset of galaxies
in the SDSS DR7 whose extinction-corrected  $r$-band Petrosian magnitudes 
are in the range 14.0 $\le m_r \le$ 17.77.  For all galaxies, we
compute local environmental densities, $\Sigma_n$:

\begin{equation}
\Sigma_n = \frac{n}{\pi d_n^2},
\end{equation}

where $d_n$ is the distance in Mpc to the $n^{th}$ nearest neighbour
within $\pm$1000 \kms.  Normalized densities, $\delta_n$,
are computed relative to the median $\Sigma_n$ within  
a redshift slice $\pm$ 0.01.  We further restrict the parent sample
to have a redshift range of 0.01 $ \le z \le$ 0.1.  The lower redshift
cut ensures cosmological values and the upper redshift cut ensures
H10 catalog completeness for LIRGs.  From this parent sample, close
galaxy pairs are selected that fulfill the following criteria:

\begin{enumerate}

\item  A projected physical separation \rp\ $\le$ 80 \hkpc.

\item  A line of sight velocity difference \dv\ $\le$300 \kms, in order
to minimize chance projections.

\item  A stellar mass ratio of 0.1 $\le$ M$_1$/M$_2$ $\le$ 10, as determined
by Mendel et al. (2013)\footnote{In this paper, we consider the mass ratio
to be a property of an individual galaxy, rather than the pair as a whole.
That is, the mass ratio is the mass of a galaxy relative to its companion.
This approach allows us to distinguish between the more (M$_1$/M$_2 > 1$) 
and less  (M$_1$/M$_2 < 1$) massive galaxy in a pair.}.

\item A normalized local density estimator $\delta_5 <$ 75,
i.e. the local density must be no more than 75 times more than the
median density in a redshift slice of $\pm$0.01 around the galaxy.
This criterion is imposed so that an adequate number of controls
can be matched to the pairs sample, see below.

\end{enumerate}

The pair selection described above is very similar to that which we
have used in previous papers (e.g. Patton et al. 2011), but with
recently re-calculated (bulge+disk) masses from Mendel et al. (2013).
As shown in Simard et al. (2011) and Patton et al. (2011), the SDSS
pipeline photometry, which is used to determine masses through
spectral energy distribution (SED) fitting, can be seriously affected
by the presence of a close companion.  Mendel et al. (2013) have
performed a thorough re-assessment of masses for the full
spectroscopic SDSS DR7 using the improved (contamination-corrected)
photometry of Simard et al. (2011).  The star formation rates used in
this paper are the `total' (aperture-corrected) values determined from
the SDSS spectra (Brinchmann et al. 2004; Salim et al. 2007).  The
star formation rates are determined through a Bayesian fitting
procedure that uses spectral energy distribution models with a range
of input parameters.  For star-forming galaxies, the SFRs are
essentially constrained using the strong Balmer emission lines, which
agree well with UV measurements (Brinchmann et al. 2004; Salim et
al. 2007).  Aperture corrections are performed based on a colour
dependent term of the light outside of the fibre region, as described
in Brinchmann et al. (2004).

The final step in the compilation of the pairs sample is a correction
for incompleteness at small angular separations due to fibre
collisions.  The relative excess of galaxies at wider separations
(where the completeness rapidly increases) can be corrected by culling
random pairs from the spectroscopic sample.  This culling procedure
randomly excludes 67.5\% of pairs at angular separations $\theta >$ 55
arcseconds and has been shown in our previous work to effectively
remove any bias in pair properties as a function of projected separation
(e.g. Ellison et al. 2008, 2011b; Scudder et al. 2012).  Of the 9397
galaxies in a pair which fulfill all of our selection criteria and
pass the cull, 400 are matched in the H10 catalog, which associates
the closest SDSS counterpart to the IRAS detection.  

Since the size of the IRAS beam is much larger than the galaxy-galaxy
separations in our pairs, the \LIR\ recorded in the H10 catalog is the
sum of the two galaxies.  We discuss in Section \ref{lirg_frac_sec}
how we account for this in terms of classifying individual galaxies as
LIRGs.  More fundamentally, we can ask whether the galaxy-galaxy
blending of IR flux is systematic for all the pairs in our sample, or
whether some pairs will be resolved by the IRAS beam, introducing a
separation dependent effect. The size of the IRAS beam is $1\times$5
arcmin, 1$\times$5 arcmin, 2$\times$5 arcmin and 4$\times$5 arcmin at
12, 25, 60 and 100 $\mu$m respectively. Most (95\%) of our pair
galaxies rely solely on the 60 $\mu$m flux, so we will consider the
effective IRAS beam size for the determination of \LIR\ to be
2$\times$5 arcmin.  97.6\% of our pairs sample have angular
separations less than 2 arcminutes and 99.6\% have separations less
than 3 arcminutes.  Therefore, for the vast majority of our pairs the
total IR flux of the two galaxies is in the single beam.  The \LIR\
measurements therefore consistently include the IR flux from both
members of a pair\footnote{More recent IR satellites offer an improved
spatial resolution, ranging from a few, to a few tens of arcseconds.
However, without a careful treatment of deblending, the photometry of
close pairs (\rp\ $<$ 20 \hkpc) can be still severely affected by near
neighbour contamination in imaging at arcsecond resolution
(e.g. Simard et al. 2011; Patton et al 2011).  It is therefore more
straighforward to deal with the fully blended IRAS images, than with
images that exhibit a `sliding scale' of contamination as a function
of projected separation.}.

\begin{figure}
\centerline{\rotatebox{0}{\resizebox{8cm}{!}
{\includegraphics{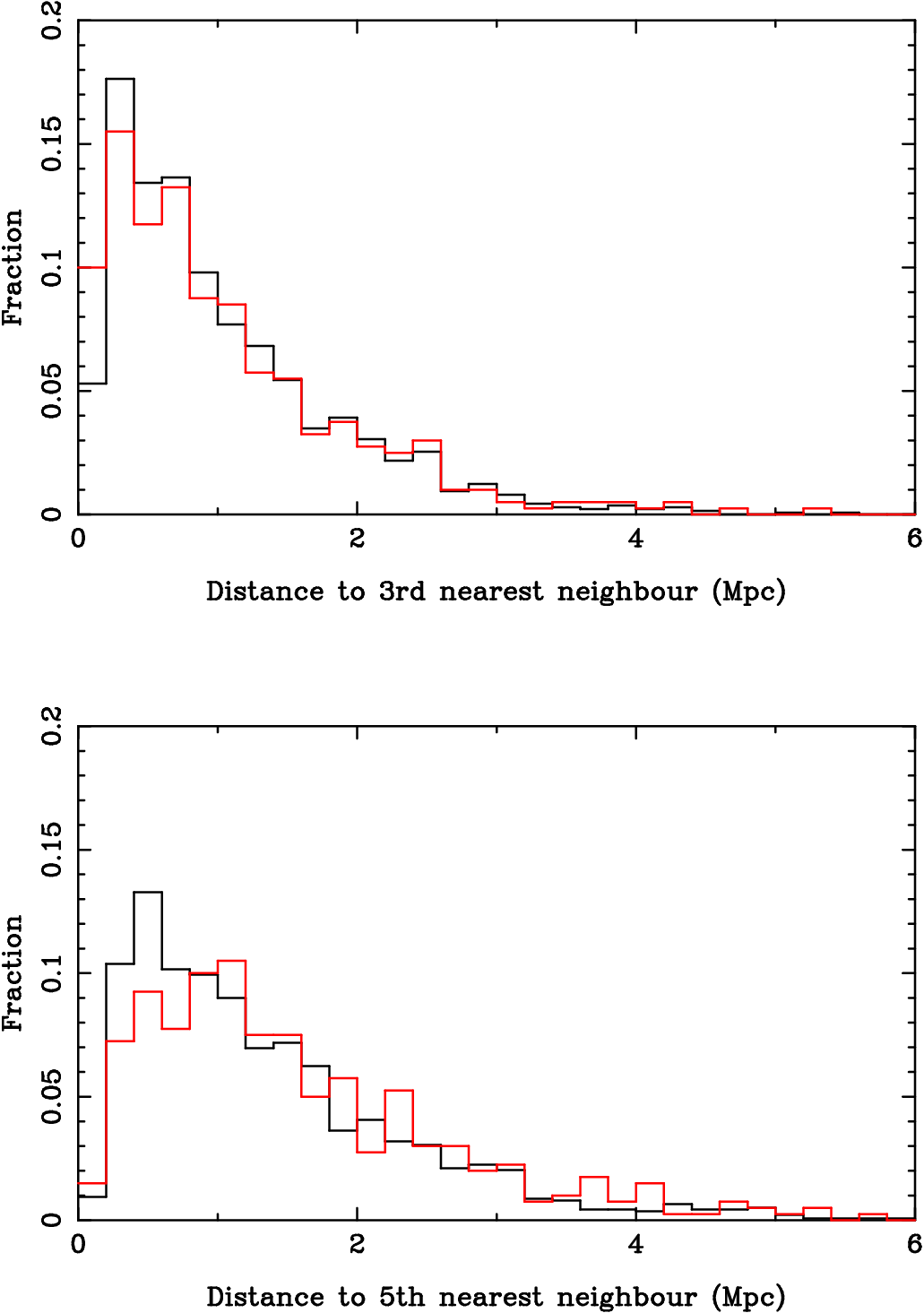}}}}
\caption{\label{neighbours} Distances to the 3rd and 5th nearest neighbour
(top and bottom panels respectively) for pairs (red) and control (black)
for galaxies that have been matched to the Hwang et al. (2010) IRAS catalog.
For both density metrics there is excellent agreement between the pair and
control distributions.}
\end{figure}

\subsection{Control sample}

\begin{figure}
\centerline{\rotatebox{270}{\resizebox{6cm}{!}
{\includegraphics{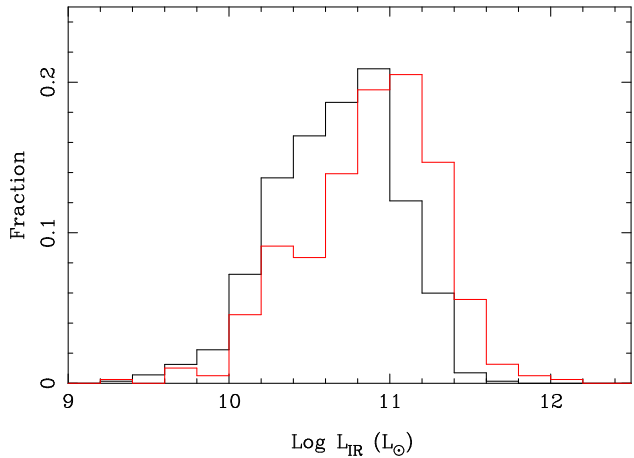}}}}
\caption{\label{pair_lir} \LIR\ distribution for our control sample
(black) and  pairs (red). Galaxies with a close companion have systematically
higher IRAS IR luminosities.}
\end{figure}

A matched control sample is an important component of understanding
the differential changes that occur in galaxies during the merger
process (e.g. Perez et al. 2009).  The sample of galaxies from which
we construct our control is initially all galaxies which have no
close spectroscopic companion within a projected separation of 80 \hkpc\
and with relative velocity less than 10,000 \kms.  However, due to the
spectroscopic incompleteness of the SDSS, this `control pool' will
still contain some truly interacting galaxy pairs for which only one
(or neither) of the galaxies has a measured redshift.  In order to
remove contaminating mergers from the control pool, we make use of
the visual classifications presented by the Galaxy Zoo project
(Lintott et al. 2008).  Darg et al. (2010b) describe the identification
of a merger sample from the Galazy Zoo, where they use an initial cut
of the weighted merger vote fraction, $f_m > 0.4$, followed-up by
further visual confirmation.  Although this vote fraction cut leads to 
a relatively 
pure merger sample, and allows a practical starting point for visual
confirmation, Darg et al. (2010b) note that the first signs of visible
disturbance can occur in galaxies with $f_m$ as low as 0.2.  Moreover,
our own visual inspection of low $f_m$ galaxies reveals that many galaxies
with $f_m \sim 0.1$ may not be strongly disturbed, but can still have a 
close companion (indeed, we have close pairs in our spectroscopic sample with 
similarly low $f_m$), although the companion tends to be faint relative 
to the main 
galaxy.  We hence conservatively exclude all galaxies with a non-zero
$f_m$ from our control pool, which reduces its size from 594,645 to
421,936 galaxies.

The control sample used in this work is matched
to the pairs sample simultaneously in mass, $z$ and environment (local 
normalized galaxy
density, $\delta_5$).  In previous work, we have taken an iterative
approach of matching the best control galaxy without replacement until
the control distributions are no longer consistent with the pairs
(e.g. Scudder et al. 2012).  This approach typically yields 10--12
matches per pair galaxy; a similar number is achieved for large
samples that have their control samples only matched in mass and
redshift (e.g. Patton et al. 2011).  However, for a large pairs
sample, matching on three parameters yields only a small number (1 or
2) of controls before the statistical comparison fails.  The failure
is usually in the distribution of densities; a small number of pairs
with atypically high local densities have few possible matches amongst
the controls.  We therefore include a maximum density threshold in the
pair selection criteria (Section \ref{pairs_sec}) of 75 times the
local median.  This excludes only $\sim$ 2.5\% of pairs\footnote{As we
show in Section \ref{vel_sec}, most IRAS detected galaxies are in
relatively low density environments.}, but allows us to match 4
controls per pair galaxy.

The environmental matching is particularly important for the present
study, due to the large size of the IRAS beam.  Even for our control
galaxies there may be `contamination' (additional IR flux) in the
measured \LIR\ from nearby galaxies.  However, for a given
environmental density, we expect the same contamination for the pairs
and control.  At $z=0.1$ the largest dimension of the IRAS beam (5
arcmin) corresponds to more than 0.5 Mpc. Since pairs tend to be in
moderately over-dense regions such as groups (Barton et al. 2007), without
environmental matching there may be additional IR flux from nearby
galaxies (the contribution from the companion is dealt with
separately, see Section \ref{lirg_frac_sec} below).  We can use our
measurements of environmental density to gain some intuition over how
well the density matching will perform on the scale of the IRAS beam.
Figure \ref{neighbours} shows the
distribution of distances to the 3rd and 5th nearest neighbours for
the pair and control galaxies that have been matched to the H10
catalog.  Although one of the parameters in the control matching
process was the locally normalized density, $\delta_5$, 
we do not necessarily
expect the pair and control histograms in Figure \ref{neighbours} to
be perfectly aligned.  This is because only galaxies with IRAS
counterparts are included, for both the pair and control distributions.
That is, whilst the matching is done independently of IRAS detections,
only those in the IRAS sample are shown in Figure \ref{neighbours}.
Nonetheless, in Fig. \ref{neighbours} there is good agreement between
the distributions of $d_5$ in the pairs and control.  There is
also an excellent agreement between the distribution of the 3rd
nearest neighbour distances of pairs and controls, as shown in the top panel of
Figure \ref{neighbours} (where, for a pair, its nearest neighbour is
the companion).  The environmental matching should therefore largely
mitigate the issue of multiple galaxies within the IRAS beam.

 Figure \ref{pair_lir} shows the \LIR\ distribution of our pairs and
control samples.  The pairs are, on average, slightly more IR luminous
than the full H10 sample, with a typical log (\LIR/\lsun) $\sim 11$.
Indeed, of the 400 paired galaxies which are matched with H10, 172 are
classified as LIRGs (40\%) and 1 as a ULIRG.  It
can already be seen from Figure \ref{pair_lir} that the pairs have
higher average \LIR\ compared to their mass, redshift and environment
matched controls.

\section{Results}

\subsection{The luminous infra-red galaxy fraction in close pairs}\label{lirg_frac_sec}

\begin{figure}
\centerline{\rotatebox{0}{\resizebox{8cm}{!}
{\includegraphics{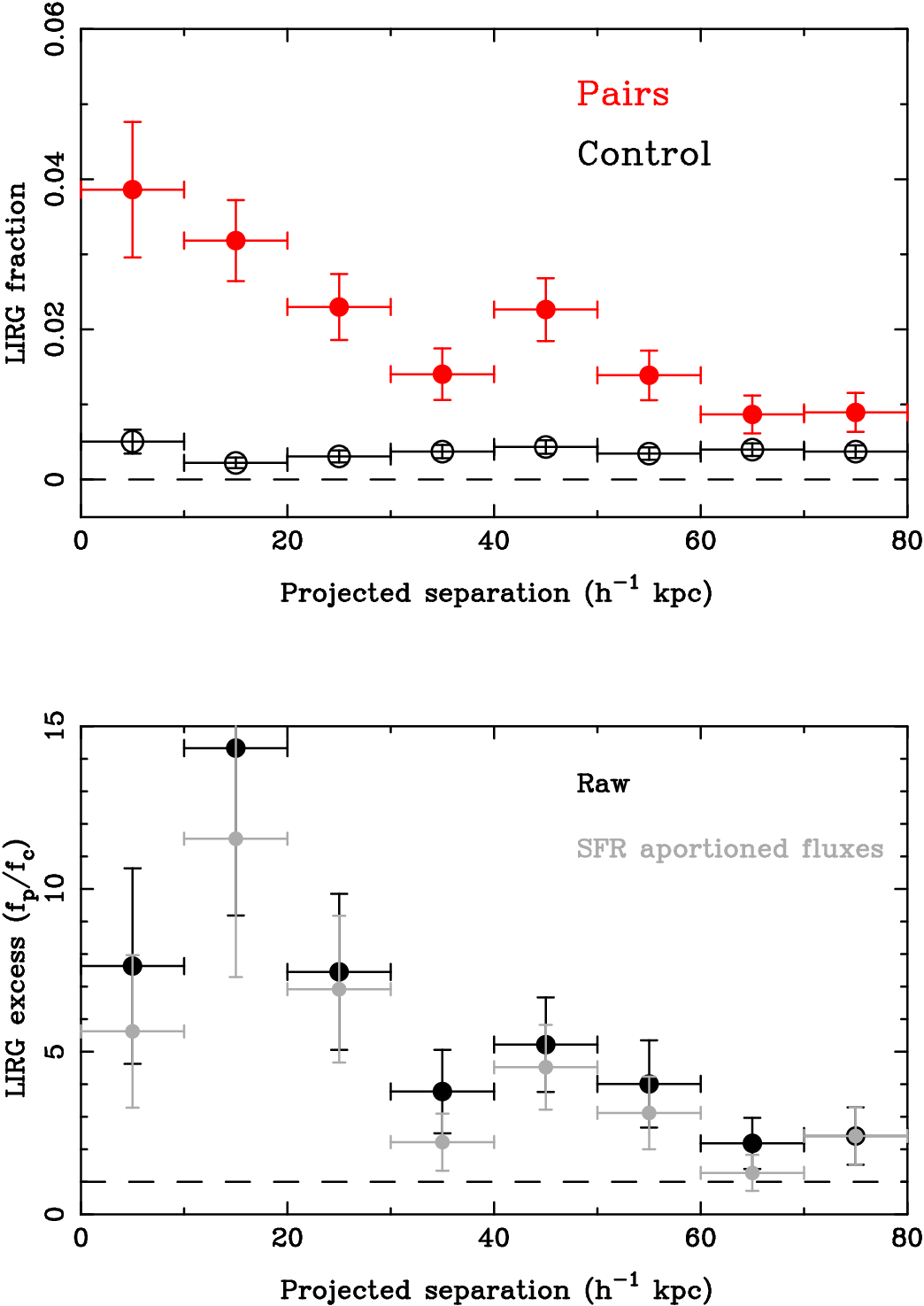}}}}
\caption{\label{lirg_frac} Top panel: Fraction of galaxies in
pairs (red) and control (open black circles) that are classified as
LIRGs (\LIR\ $\ge 10^{11}$ \lsun) as a function of projected
separation.  Lower panel: The ratio of LIRG fractions in
pairs-to-control as a function of projected separation.  
The grey shaded circles have been corrected for the IR flux contribution
from the companion galaxy.  This figure reveals that the fraction of LIRGs
is elevated in close pairs at all separations in our sample, averaging 5--10
times the control fraction at small \rp.}
\end{figure}

We begin our investigation of the IR properties of pairs by
quantifying in Figure \ref{lirg_frac} the fraction of LIRGs in our
pairs sample as a function of projected separation.  The LIRG fraction
in each \rp\ bin is calculated relative to the total number of
pair galaxies in that \rp\ range, regardless of IRAS detection.  There is a
steady increase in the LIRG fraction in the pairs as the projected separation
decreases, indicating that their high \LIR\ is associated with the
interaction.  In addition to the general increase in the LIRG fraction
amongst the pairs sample, it is of interest to estimate the relative
LIRG frequency compared to the control sample.  The lower panel of
Fig. \ref{lirg_frac} shows that the LIRG fraction in pairs is enhanced
relative to the control sample at small separations ($r_p < 20$ \hkpc)
by a factor of $\sim$
5 -- 10, and remains enhanced at the highest separations of the sample
(80 \hkpc) by up to a factor of $\sim$ 2 -- 3.  
  
It is important to consider whether the LIRG fraction enhancement is
just extra IR flux moving into the beam as galaxies get closer
together.  Similarly, one might question whether the LIRG fraction
plateau at wide separations is caused by residual environmental
effects not accounted for in our control sample matching.  As argued
in Section \ref{pairs_sec}, for all but the widest pairs ($\sim$2\%),
both galaxies are in the beam, so that the trend of increasing LIRG
fraction as a function of \rp\ can not be due to a gradual trend of
beam confusion.  Moreover, since the projected separations of
this minority of partially resolved galaxies are all $r_p > 30$ \hkpc,
separation dependent beam confusion/resolution can not
explain the rise in the LIRG fraction at small projected separations.
As a further test, we construct a sample of projected (in redshift
space) pairs that we do not expect to show any signs of interaction, but are
nonetheless close on the sky.  We have previously used such projected
samples to test for photometric contamination effects (e.g. Patton et
al. 2011).  The projected sample is selected identically to the main
pairs sample, with the exception that the relative velocity criterion
is now 300 $<$ \dv\ $<$ 5000 \kms.  We first confirm that the SFRs of
the projected sample do not depend on \rp.  Next, the LIRG fraction of
the projected pairs is calculated and found to be flat with \rp.  Had
the IRAS beam been gradually missing flux at larger \rp, then the LIRG
fraction of the projected pairs would have decreased with increasing
separation.  We therefore conclude that the LIRG fraction of the 
\dv\ $\le$ 300 \kms\ pairs is truly declining with \rp, and the effect
shown in Fig. \ref{lirg_frac} is not a beam confusion effect.

It should be noted that although our \LIR\ measurements are the
combined fluxes from two galaxies, such blending is also frequently
present in other LIRG and ULIRG samples (e.g. Howell et al. 2010).
Nonetheless, it is of interest to attempt a decomposition of the \LIR\
between the two galaxy companions.  As argued above, the environmental
matching should largely account for this; subtracting the estimated IR
flux from the companion is a very conservative correction.  We assume
that the \LIR\ scales as the SFR, a relationship that is
well-established in the literature and present in our control sample.
The \LIR\ correction is therefore applied by dividing the total \LIR\
between the two galaxies relative to their respective SFRs.  Depending
on the initial \LIR\ in the H10 catalog, and the relative SFRs, this
can either increase or decrease the number of LIRGs in each \rp\ bin
in Figure \ref{lirg_frac}.  For example, if a pair galaxy with log
(\LIR/\lsun) = 11.1 that was originally classified as a LIRG has a
companion with an identical SFR, the two galaxies will both be
classified as non-LIRGs after the \LIR\ has been equally divided.  On
the other hand, a pair galaxy with log (\LIR/\lsun) = 11.9 would have
originally been classified as a LIRG, but since no IRAS detection is
associated with its companion (due to the large IRAS beam), only 1
LIRG is counted for the pair.  However, if the SFRs of the two
galaxies are equal, the equal division of the \LIR\ will yield two
LIRGs.  The overall result of the flux aportioning is shown by the
grey shaded points in Figure \ref{lirg_frac}.  The SFR-corrected LIRG
fractions generally decrease, but there is still a trend with \rp\
and an enhancement at all separations over the control by a factor of
2--10.

\subsection{Pairwise properties of LIRGs}

\begin{figure}
\centerline{\rotatebox{0}{\resizebox{8cm}{!}
{\includegraphics{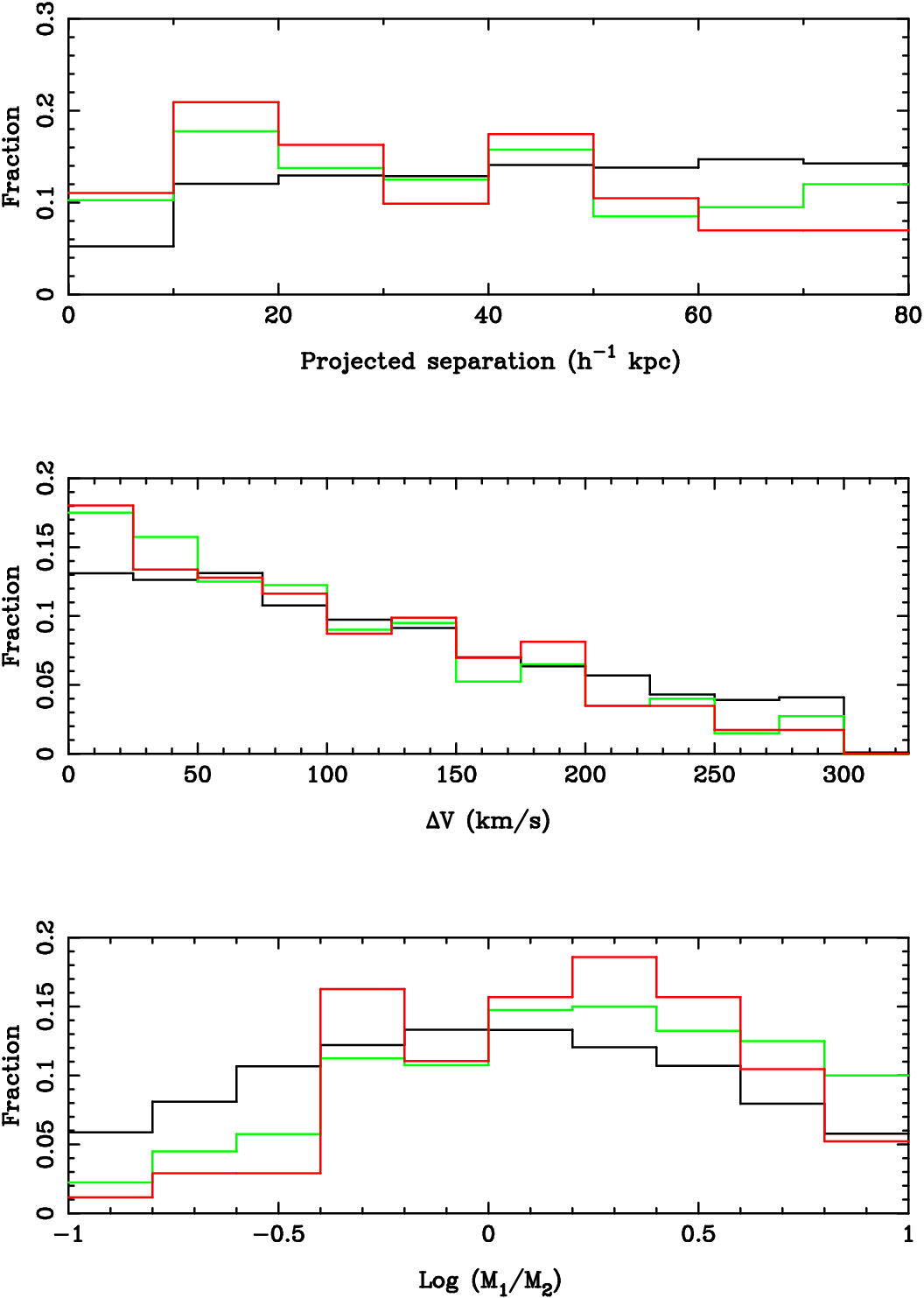}}}}
\caption{\label{props} Pairwise properties for the full pairs sample
(9397 galaxies, black), those detected by IRAS (400 galaxies, green)
and those classified as LIRGs (172 galaxies, red). Mass ratios are
log (M$_1$/M$_2$) > 0 for the more massive companion and log (M$_1$/M$_2$) 
< 0 for the lower mass companion in a pair.  IRAS detections and
LIRGs are biased towards more massive galaxies (see Figure \ref{individual}, 
hence the asymmetry in
the red and green histograms, relative to the full sample (black).}
\end{figure}

\begin{figure}
\centerline{\rotatebox{0}{\resizebox{8cm}{!}
{\includegraphics{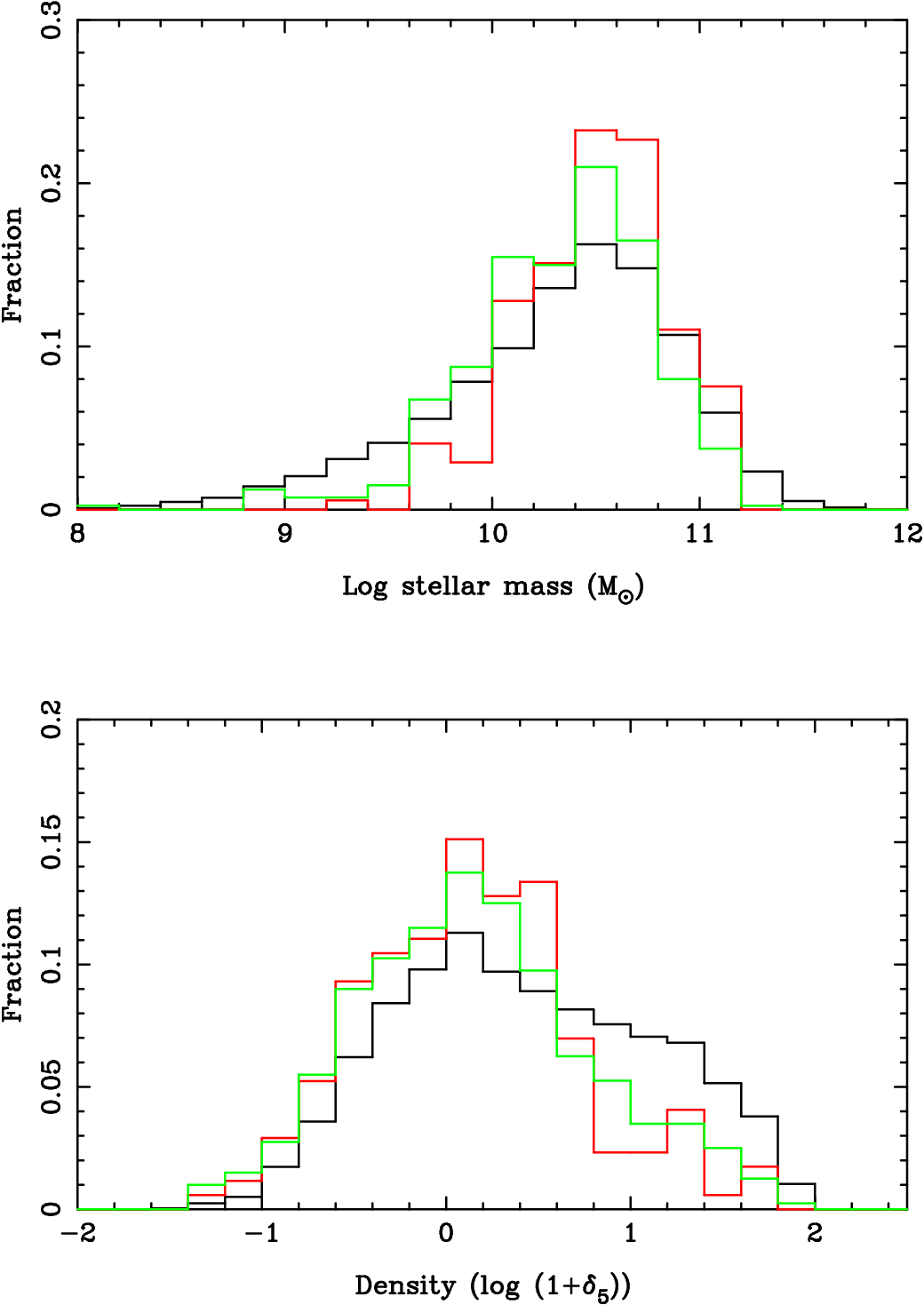}}}}
\caption{\label{individual} Galaxy masses and local galaxy densities
(the environmental metric used in this work) for the full pairs sample 
(9397 galaxies,
black), those detected by IRAS (400 galaxies, green) and those
classified as LIRGs (172 galaxies, red). LIRGs are preferentially
found in lower density environments relative to the full sample, but
in a similar range of densities to other IR detected (star-forming)
galaxies.}
\end{figure}

We next investigate whether there are certain pair configurations
that are particularly conducive to producing LIRGs.  Figure
\ref{props} compares the properties of the pairs that are LIRGS,
relative to the full pairs sample that fulfill the criteria listed in
Section \ref{sample_sec}.  The full pairs sample includes
both actively star-forming and quiescent galaxies.  Since we expect
the distribution of fundamental properties of these two categories
to differ, we also compare the LIRG pairs with the pairs with IRAS
detections, but with log (\LIR/\lsun) $<$ 11, which are therefore
star-forming, but presumably at lower rates than the LIRGs, or with a lower
dust content.

\subsubsection{Projected separation}

The top panel of Figure \ref{props} shows that the LIRG pairs (red
histogram) tend to have smaller projected separations than the full
pairs sample, as expected from Figure \ref{lirg_frac}.  A
Kolmogorov-Smirnov (KS) test yields a probability of 1.6$\times
10^{-6}$ that the full pairs sample (black histogram) is drawn from the
same \rp\ distribution as the LIRGs (red histogram).  This is also
consistent with the finding by Hwang et al. (2010) that higher \LIR\
galaxies tend to have closer nearest neighbours.  The bias towards low
\rp\ is also seen in all the IRAS detected pairs (green histogram),
where the LIRG and IRAS detected \rp\ distributions are
indistinguishable in a KS test.  This indicates that SFR enhancements
contribute over a range of IR luminosities.  There is also a hint of a
\textit{deficit} of LIRG pairs relative to the full pairs sample at
wide separations, \rp\ $>$ 50 \hkpc.  Recall that the LIRG fraction
relative to the control is still enhanced out to at least \rp\ = 80
\hkpc\ (Figure \ref{lirg_frac}).  The relative paucity of LIRG pairs
relative to non-LIRG pairs simply indicates that the SFR enhancements
are likely to be more modest at these wide separations, as
demonstrated explicitly in Scudder et al.  (2012).

\subsubsection{Relative velocity}\label{vel_sec}

The middle panel of Figure \ref{props} shows that the LIRG pairs may
be slightly skewed towards lower \dv\ values than the full pairs
sample, although a KS test does not yield any significant difference
between the \dv\ distributions of any of the samples.  Nonetheless,
a bias towards smaller \dv\ would be consistent with previous
observational results that have found that the SFR of close galaxy
pairs increases with decreasing \dv\ (e.g. Lambas et al. 2003; Nikolic
et al. 2004; Alonso et al. 2004, 2006).  However, the measurement
of SFR dependence on \dv\ is complicated by an underlying correlation
between \dv\ and local galaxy density (Ellison et al. 2010), which
means that \dv\ is itself a coarse metric of environment.  In turn,
there is also a well established SFR-density relation in the local
universe (e.g. Gomez et al. 2003), so that a bias towards small \dv\
values in LIRGs could be associated with higher intrinsic SFRs in
lower density environments.  In the bottom panel of Figure
\ref{individual} we show that the LIRG pairs are certainly biased
towards lower density environments than the full pairs sample.
However, the density distribution of LIRGs is similar to that of other
IRAS detected star-forming pairs (green histogram in Figure
\ref{individual}) (see also Hwang et al. 2010).  The environments of
the full pairs sample approximates a double log-normal density
distribution, where the high density population is apparently
dominated by quiescent galaxies (Baldry et al. 2006).

\subsubsection{Mass ratio}\label{massr_sec}

Our complete pairs sample is, by design, symmetric in mass ratio
(black histogram in Figure \ref{props}).  However, the pairs that are
identified as LIRGs are biased towards high mass ratios (red histogram
in Figure \ref{props}), i.e. the LIRG galaxies in pairs tend to be
more massive than their companions.  A KS test yields a 2.3$\times
10^{-5}$ probability that the LIRGs are drawn from the same distribution of
mass ratios as the full pairs sample.  There are very few LIRG
galaxies that have masses less than a factor of 3 below that of their
companion.  The magnitude limited nature of the SDSS survey means that
high mass ratio galaxies are amongst the most massive in the sample.
The tendency for LIRG pairs to have high mass ratios is therefore
likely to be a selection effect.  Galaxies with higher masses (and
intrinsically higher SFRs) require a smaller SFR boost to achieve LIRG
status (a topic which is discussed in more detail in Section
\ref{how2_sec}).  Indeed, relatively few of the LIRGs in either our pair or
control samples have stellar masses log (M$_{\star}$/\msun) $< 10$ (Figure
\ref{individual}).  However, the lower panel of Figure \ref{props}
hints that the LIRG pairs have a narrower distribution of mass ratio
than the IRAS detected pairs (although a KS test is inconclusive), a
result which could not be explained by mass bias.  Simulations have
previously predicted that the most effective SFR boost occurs within
almost equal mass mergers (Cox et al. 2008).  Observational works that
have measured direct enhancements in the SFRs of close pairs support the
dominant role of major mergers in producing the strongest starbursts
(Woods et al. 2006; Ellison et al. 2008; Scudder et al. 2012; Lambas
et al. 2012).  The largest IR luminosities may therefore be associated
with galaxies that are both massive (and therefore have relatively
high SFRs to start with) and are engaged in major mergers (Hwang et
al. 2010).  Indeed, massive galaxies engaged in major mergers seem to
dominate the local ULIRG population (Dasyra et al. 2006).

\begin{figure}
\centerline{\rotatebox{270}{\resizebox{6cm}{!}
{\includegraphics{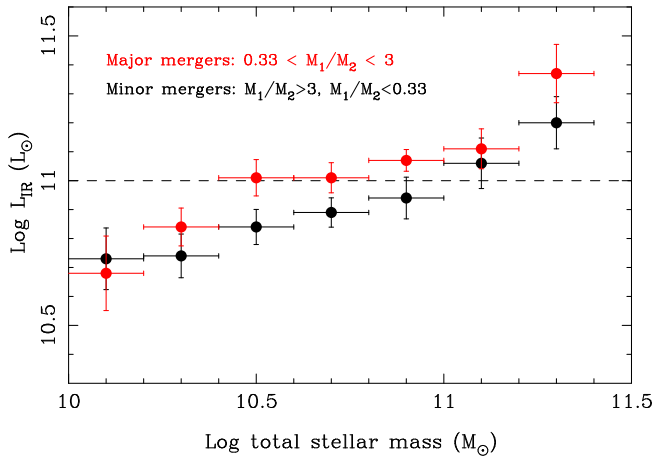}}}}
\caption{\label{mtot} \LIR\ as a function of the total pair mass
($M_1+M_2$).    Red points
indicate major pairs (0.33 $< M_1/M_2<$ 3) and black points show minor
pairs up to the maximum mass ratio of our sample, 1:10. The horizontal
dashed line is drawn at log (\LIR/\lsun) = 11.  The offset to higher
\LIR\ at fixed total stellar mass in major pairs indicates higher
levels of triggered star formation.}
\end{figure}

To further investigate the dependence of LIRG classification on mass
and mass ratio, in Figure \ref{mtot} we plot the median \LIR\ as a
function of the total stellar mass of the pair.  We distinguish
the galaxy pairs by their relative masses, in order to test whether,
at fixed total mass, the mass ratio has any impact on the total \LIR\
(i.e. SFR).  Galaxies with masses within a factor of three of each
other are considered to be equivalent to major mergers (shown in red
in Figure \ref{mtot}); all others (up to a maximum mass contrast of a
factor of 10) are designated as minor mergers (black points).
Figure \ref{mtot} shows that \LIR\ increases with the 
total stellar mass of the pair, as expected from the correlations of mass with 
SFR and \LIR.  However,  it can also be seen that at fixed total
stellar mass, the median \LIR\ is systematically higher for major mergers.
The systematic offset to higher \LIR\ at fixed total stellar mass in
major mergers supports higher rates of dusty star formation in
approximately equal mass encounters. 

\subsection{The role of AGN}\label{agn_sec}

In the previous sections, we have tacitly assumed that the main driver
of the LIRG phase in our pairs sample is the enhancement in the SFRs
that is a well-established result of galaxy interactions.  In moderate
luminosity LIRGs, the \LIR\ indeed appears to be dominated by star
formation (e.g. Alonso-Herrero et al. 2010).  However, AGN can also
contribute to the IR luminosity.  Petric et al. (2011) confirm the
modest contribution from AGN in the majority of LIRGs from a study of
high ionization lines in the mid-IR spectra of 248 nearby LIRGs.  They
find that $\sim$ 20\% of LIRGs contain an AGN, and that the AGN is the
dominant contributor of IR flux in only 10\% of LIRGs. The relative
importance of AGN to the IR flux increases with \LIR, so that the
ULIRGs exhibit both high AGN fractions, and higher contribution from
AGN within a given galaxy (e.g. Yuan et al. 2010 and references
therein).  In isolated star-forming galaxies, there is a well-known
increase in the AGN (as classified from optical emission line ratios)
fraction with stellar mass.  Since the SFR also increases with stellar
mass, the connection between AGN incidence and \LIR\ is not
too surprising.  However, there seems to be an additional connection
between the AGN incidence in IR luminous galaxies and mergers.  The
AGN-dominated LIRGs (Petric et al. 2011) and ULIRGs (Yuan et al.
2010) seem to preferentially occur at late merger stages.  Combined
with the observation that our pairs show an enhanced AGN fraction
towards small separations (Ellison et al. 2011b), it is germane to
consider whether the increase in LIRG fraction seen in Figure
\ref{lirg_frac} is due to an increasingly dominant contribution from
AGN.

\begin{figure}
\centerline{\rotatebox{0}{\resizebox{8cm}{!}
{\includegraphics{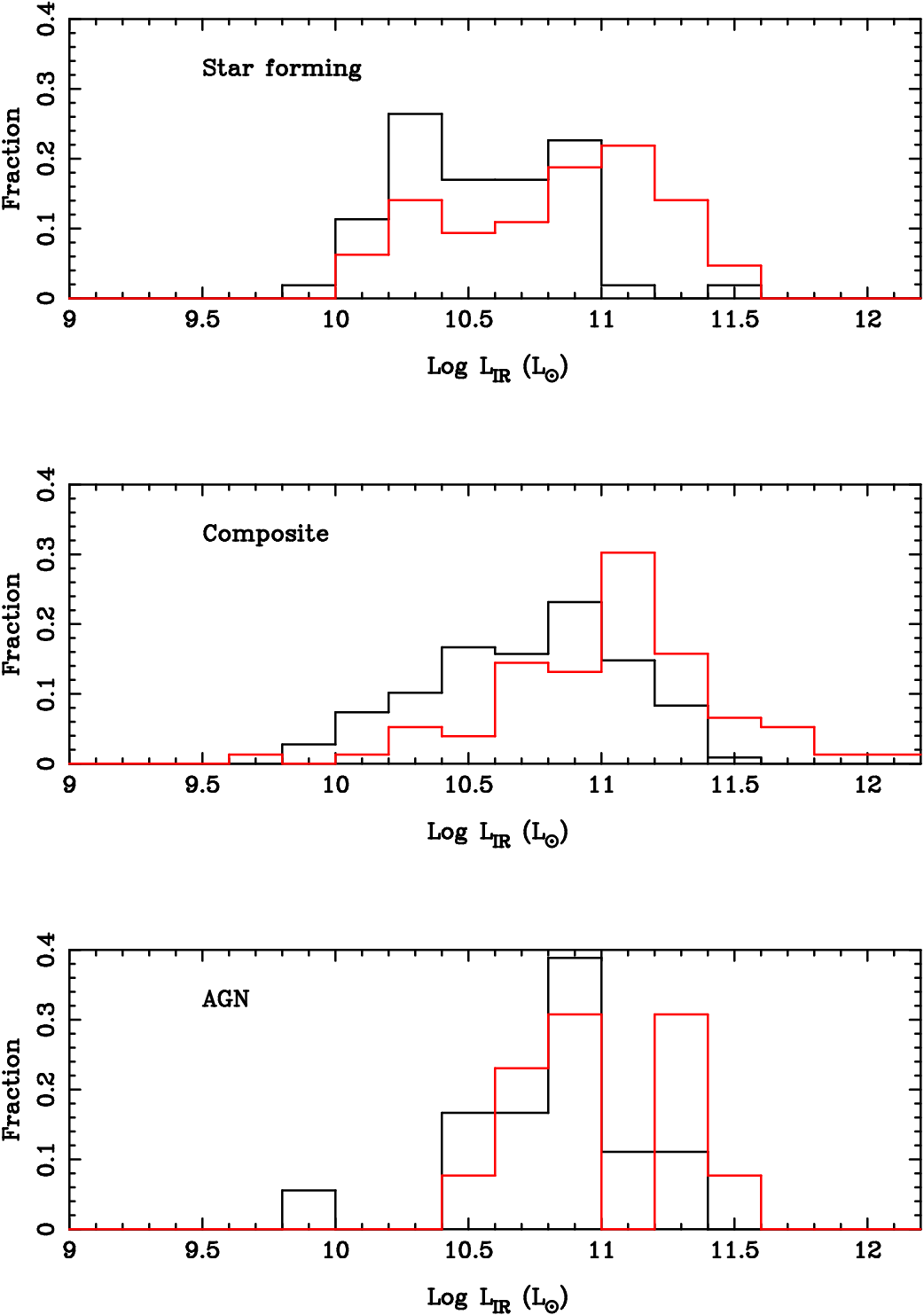}}}}
\caption{\label{lir_hist_AGN} \LIR\ distributions for star-forming
(top panel), composite (middle panel) and AGN (lower panel) galaxies
in our close pairs (\rp\ $\le30$ \hkpc, red) and control (black)
samples. The enhancement of \LIR\ in pairs relative to control is seen
in the star-forming and composite galaxies, but not in the AGN.  }
\end{figure}

Figure \ref{lir_hist_AGN} shows the \LIR\ distributions for pairs and
control (red and black histograms, respectively) in star-forming,
composite and AGN galaxies.  The three categories are defined by
combining the classifications of Kewley et al. (2001) and Stasinska et
al (2006), and are made for galaxies with emission line
signal-to-noise ratios S/N$\ge$3.  For this figure, we consider only
pairs (and their controls) with \rp\ $\le 30$ \hkpc, where differences in
SFR, AGN and LIRG fraction (relative to the control sample)
are most significant.  Recall that we are
only complete for log (\LIR/\lsun) $>$11, but this incompleteness
applies equally to all our galaxy classes.  It is also important to
keep in mind that amongst the emission line galaxies in our sample,
there is a higher AGN fraction at higher stellar masses, where the
SFRs (and hence \LIR) are, on average, higher.  Figure
\ref{lir_hist_AGN} shows a shift to higher \LIR\ for AGN versus star
forming galaxies, e.g. by comparing the control samples for the two
categories (black histograms in lower and upper panels respectively).
Such a shift is expected based on the mass distributions of the AGN
and star-forming galaxies, as described above.  Figure
\ref{lir_hist_AGN} also shows that the pairs (red) have systematically
higher \LIR\ than the mass-matched controls, for both star-forming and
composite classifications.  The mass-matching means that the \LIR\
differences between the pair and control galaxies is driven by the
interaction, rather than a mass bias.  However, there is no
statistically significant difference between the \LIR\ distributions
of the pair and control AGN samples (Figure \ref{lir_hist_AGN}, lower
panel).  A KS test yields a $\sim$ 50\% probability
(i.e. inconclusive) that the two samples are drawn from different
populations.  We conclude that the increase in LIRG fraction seen in
Figure \ref{lirg_frac} is not due to an increasing contribution to the
\LIR\ from AGN, but rather is driven by star formation.

\begin{figure}
\centerline{\rotatebox{270}{\resizebox{6cm}{!}
{\includegraphics{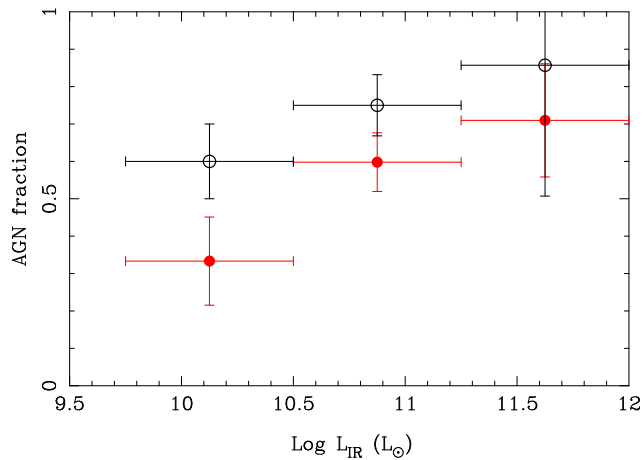}}}}
\caption{\label{AGN_frac} Fraction of galaxies in close pairs (\rp\ $\le30$ 
\hkpc, red) and control (open circles) that are classified as AGN as a 
function of \LIR.  }
\end{figure}

Having argued that the LIRG pairs are dominated by star formation, we
can now turn to the question of whether the merger contributes to the
\LIR\ (or mass) dependence of AGN incidence.  Figure \ref{AGN_frac}
shows the fraction of AGN as a function of \LIR\ for both our close 
(\rp\ $\le$ 30 \hkpc) pairs
and control sample.  Again, we recall that although our sample is only
complete to log (\LIR/\lsun) $>$ 11, the pairs and control can be compared
differentially.  In agreement with previous studies (e.g. Goto 2005; Hwang et
al. 2010; Kartaltepe et al. 2010; Yuan et al. 2010; Iwasawa et
al. 2011) the AGN fraction increases steadily in the control with
increasing \LIR. As described above, this is likely due to the
co-dependence of stellar mass on both SFR (which correlates with \LIR)
and AGN fraction, such that higher stellar mass galaxies have higher
SFRs (and hence higher \LIR) and higher AGN fractions.  Note that the
exact AGN fraction at fixed \LIR\ is highly dependent on the
classification process (e.g. different line ratio cuts in the optical,
colour classifications in the mid-IR and X-ray classification),
making a direct comparison with other works non-trivial.  Even for a
given technique, such as optical line ratio diagnostics, precise
classification is sensitive to choices such as S/N, as lower S/N cuts
will increasingly include low ionization nuclear emission line regions
(LINERs, e.g. Cid-Fernandes et al. 2010).  The AGN classification of
Stasinska et al. (2006) includes composites in which even a minor
contribution of AGN is inferred from the line ratios.  Using a
`stricter' AGN classification, such as that of Kauffmann et al. (2003)
reduces the AGN fractions by a factor of about 2, but still leaves the
trend with \LIR\ intact, yielding an AGN fraction of $\sim$ 20\% for
$10<$ log (\LIR/\lsun) $<11$ and about 50\% for the LIRGs.

Beyond the general increase of AGN fraction with \LIR, Figure
\ref{AGN_frac} hints that the AGN fraction in pairs may be slightly
lower than the control at fixed IR luminosity.  Such behaviour might
be expected if pairs increased their \LIR\ without increasing their
AGN fraction.  This is in apparent disagreement with several recent
studies that have linked mergers with AGN triggering in the local
universe (Ellison et al. 2011b; Liu et al. 2012; Koss et al. 2012).
One solution to this tension may be that the dust that is producing
the high IR luminosities in our sample is hindering the
classification of optical AGN.  A forth-coming paper which uses
multi-wavelength mid-IR diagnostics will investigate dust-obscured
AGN in close pairs.

\section{Discussion}

\subsection{Merger fractions}

It is widely accepted that the fraction of mergers increases as a
function of \LIR, with mergers representing the bulk of the ULIRG
population (e.g. Kartaltepe et al. 2010).  At LIRG luminosities, 15-50
\% of galaxies are usually classified as mergers (e.g. Zheng et
al. 2004; Melbourne et al. 2005; Wang et al. 2006; Kaviraj 2009;
Kartaltepe et al. 2010).  However, these studies all start with an IR
selected sample and classify mergers based on visual morphologies.
Our study of spectroscopically selected close pairs can offer an
alternative measure of the merger fraction as a function of \LIR.

In order to determine the fraction of mergers as a function of \LIR,
we must first define the term `merger'.  As for the AGN statistics in
the previous section, we set an upper \rp\ limit of 30 \hkpc\ as the
separation within which the strongest interaction-induced effects are
seen.  The parent pairs sample has a mass ratio range from 1:10 to
10:1.  However, at higher redshifts the magnitude limited nature of
the SDSS means that minor mergers become increasingly scarce and we
are dominated by pairs with approximately equal masses (and
magnitudes), see Section \ref{massr_sec}.  It is additionally
necessary to consider the mass completeness of our sample.  For
galaxies near the (mass) detection threshold of the SDSS, we will be
unable to detect lower mass companions.

In order to tackle the above issues of completeness, the measurement
of a merger fraction requires a careful definition of `close
companion'.  The first component of this definition is in terms of
companion proximity, both in projected and radial space.  We adopt the
criteria of a close companion to be a galaxy within 30 \hkpc\ and 300
\kms.  We also restrict our definition to major mergers, i.e. galaxies
whose companion has a mass within a factor of three of its own.  The
mass completeness of the SDSS is strongly colour dependent .  At fixed
redshift, blue galaxies can be detected to considerably lower masses
than red galaxies (e.g. van den Bosch et al. 2008; Mendel et
al. 2013), due to the flux limit in the SDSS.  At a redshift of
$z=0.1$, the SDSS is complete for both red and blue galaxies above log
(M$_{\star}$/\msun) $\sim$ 10.7 (Mendel et al. 2013).  A complete
census of all 0.33 $\le$ M$_1$/M$_2$ $\le 3$ companions would
therefore require a mass limit at $z=0.1$ of log (M$_{\star}$/\msun)
$>$ 11.2, and hence severely limit the sample for which we could
determine merger fractions.  We therefore adopt a mass limit that
only includes blue cloud galaxies (Mendel et al. 2013), and caution
that our close pair fractions are only applicable in this mass range
(at $z=0.1$, the mass limit corresponds to (M$_{\star}$/\msun) $\sim$ 10.37;
for lower redshifts, the limit will be lower).

The process for the quantification of merger fraction as a function of
\LIR\ is as follows.  For each galaxy in the SDSS spectroscopic sample
that appears in the Hwang et al. (2010) IRAS catalog, we determine
whether its stellar mass is at least 0.5 dex (a factor of three) above
the redshift dependent completeness for blue galaxies described in
Mendel et al. (2013).  Every IRAS-detected SDSS galaxy that passes
this mass criterion is then compared with the subset of our catalog of
close pairs that have \rp\ $\le$ 30 \hkpc, \dv\ $\le$ 300 \kms, and
0.33 $\le$ M$_1$/M$_2$ $\le 3$ to identify whether the galaxy has a
companion.  During the compilation of the pairs sample, pairs
with angular separations $\theta > 55$ arcseconds are randomly culled
to produce an even fibre (in)completeness as a function of separation.
Patton \& Atfield (2008) showed that the level of spectroscopic
completeness at these small separations is $\sim$ 26\%, relative to
the photometric galaxy catalog.  This completeness rate includes both
the effects of fibre collisions and the overall spectroscopic
completeness of the SDSS.  We therefore multiply our pair fractions by
a factor of 3.85.

\begin{figure}
\centerline{\rotatebox{270}{\resizebox{6cm}{!}
{\includegraphics{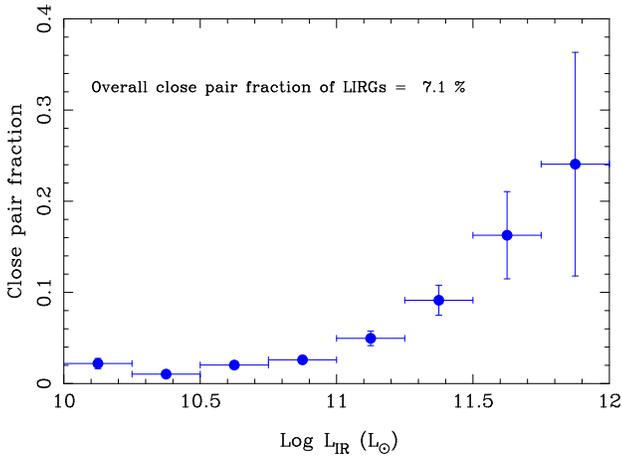}}}}
\caption{\label{merger_frac} Fraction of galaxies that have a close
companion within 30 \hkpc, 300 \kms\ and 0.33 $\le$ M$_1$/M$_2$ $\le$
3 as a function of \LIR.   All merger fractions
have been corrected for fibre collision incompleteness and the general
spectroscopic incompleteness of the SDSS.  }
\end{figure}

Figure \ref{merger_frac} shows the fraction of
galaxies in each \LIR\ bin that have a close companion within the
criteria for proximity and mass completeness described above.  
We find that below log (\LIR/\lsun) $\sim$ 11 the fraction of galaxies 
that have a close, approximately equal mass companion is approximately 
constant at $\sim$ 2\%.  Above LIRG
luminosities there is a rise in the fraction of galaxies that have a
close companion, with an overall pair fraction amongst the LIRGs of
7\%.   It is not clear why the pair fraction increases only above
log (\LIR/\lsun) $>$ 11, but a similar dependence has been previously
seen (e.g. Kartaltepe et al. 2010).  Tekola, Vaisanen \& Berlind
(2012) have shown that for galaxies with log (\LIR/\lsun) $<$ 11 there
is no correlation between IR luminosity and local galactic density.
However, they also showed that for LIRGs, there is a strong
correlation between \LIR\ and density, and suggested that the LIRG
luminosity (at low redshifts) represents a real transition in galaxy
properties.

Although the close pair fraction at LIRG luminosities that we find
is of a similar magnitude to what has been found in previous studies 
(e.g. Zheng et al. 2004; Melbourne et al. 2005; Wang et al. 2006; Kaviraj 
2009;  Kartaltepe et al., 2010), a rigourous quantitative comparison (i.e.
on the basis of absolute merger fractions) with
other works is non-trivial.  For example, we have defined our mergers
in a specific projected separation, \dv\ and mass ratio range. The
fraction of LIRGs (and SFRs of pairs in general) remain elevated
beyond our nominal cut-off of \rp\ $\le$ 30 \hkpc.  The mass ratio cut
is dictated by completeness considerations, but minor mergers can
experience significant SFR enhancements (e.g. Scudder et al.  2012)
that can result in LIRGs (Figure \ref{props}).  Indeed, Hwang et
al. (2010) find that $\sim$34\% of the LIRGs in their low redshift
sample (also a result of a cross-matching between IRAS and SDSS) are
in minor mergers.  Our close pair fraction of 7\%  might therefore
be considered to 
be a lower limit to the actual fraction of LIRGs with a companion.
Conversely, our close pairs sample will also include
galaxies that have not yet interacted (i.e. have not yet experienced a
pericentric passage), as well as some projected pairs that are not
actually destined to interact.  Both of these categories will dilute
the fraction of LIRGs associated with merger-triggered star formation.
Scudder et al. (2012) visually classified a subset of our
star-forming pairs and found only $\sim$ 55\% to show visible signs of
disturbance.  Finally, we do not include a merger timescale, or
an observability timescale.  These various caveats serve to caution
about an absolute comparison with other works, but the general
trend of a rising close pair fraction at high \LIR\ should be robust,
and the results in Figure \ref{merger_frac} demonstrate a connection
between mergers and high IR luminosities.

\subsection{Star formation rates and extinction}

\begin{figure}
\centerline{\rotatebox{0}{\resizebox{8cm}{!}
{\includegraphics{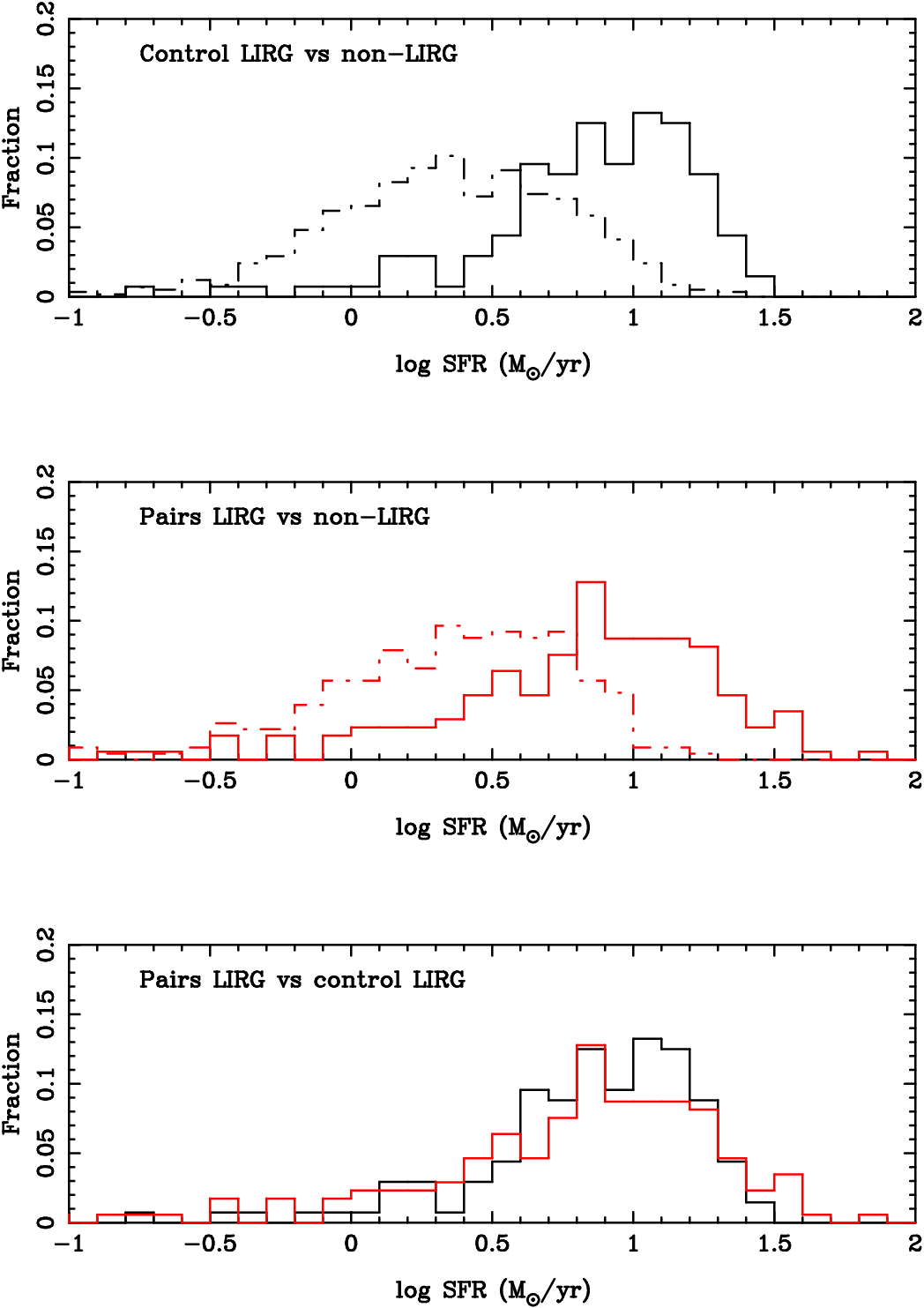}}}}
\caption{\label{sfr} Top panel: The distribution of SFRs in control
galaxies for LIRGs and non-LIRGs.  Middle panel: The distribution of
SFRs in pair galaxies for LIRGs and non-LIRGs. Bottom panel: The
distribution of SFRs in LIRG galaxies in the pair sample and
control sample. In all panels, pairs are shown in red, control in black;
LIRGs are shown with a solid line and non-LIRGs with a dot-dashed line.}
\end{figure}

\begin{figure}
\centerline{\rotatebox{0}{\resizebox{8cm}{!}
{\includegraphics{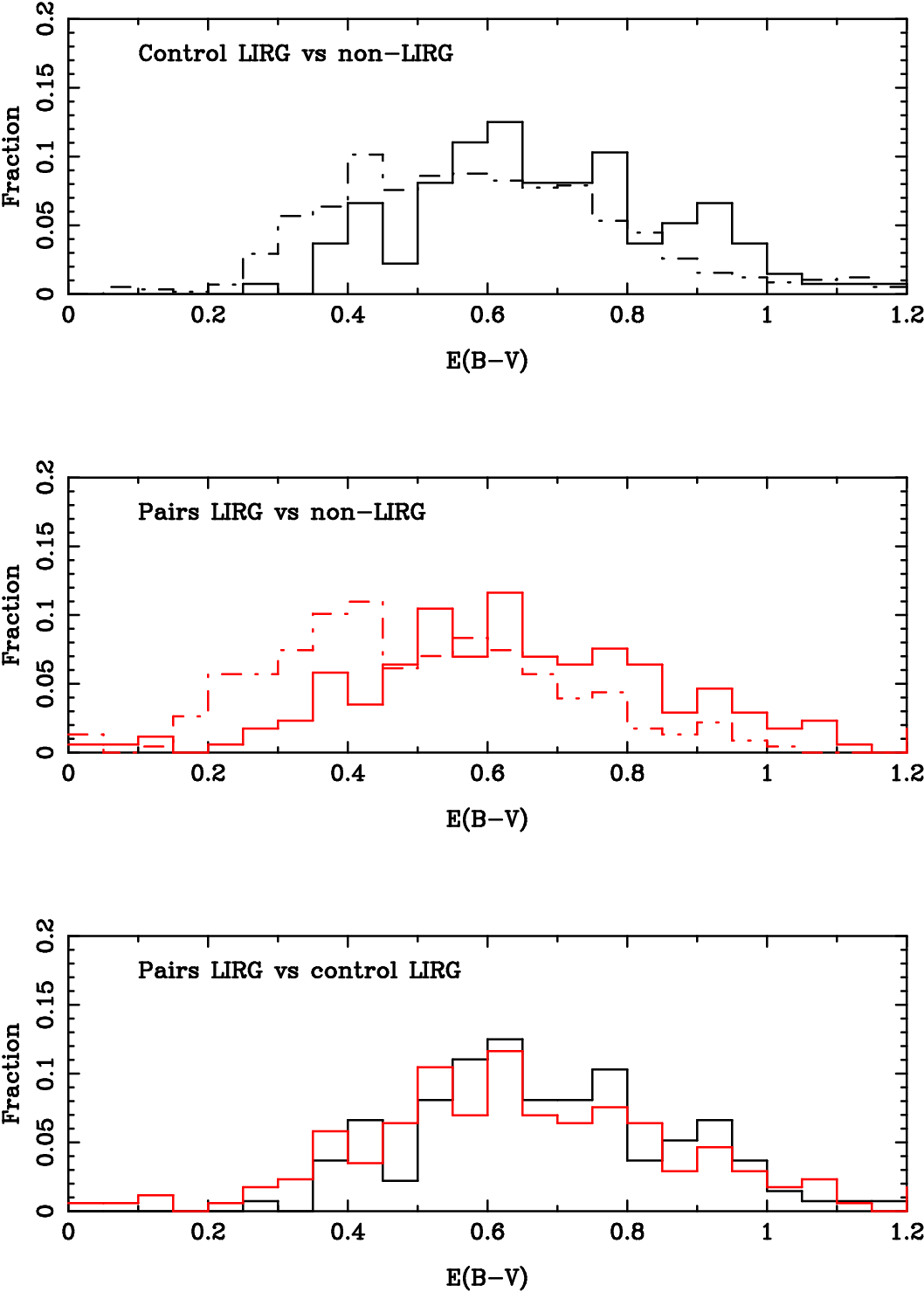}}}}
\caption{\label{ebmv} Top panel: The distribution of \ebmv\ in control
galaxies for LIRGs and non-LIRGs.  Middle panel: The distribution of
\ebmv\ in pair galaxies for LIRGs and non-LIRGs.  Bottom panel: The
distribution of \ebmv\ in LIRG galaxies in the pair sample and control
sample. In all panels, pairs are shown in red, control in black; LIRGs
are shown with a solid line and non-LIRGs with a dot-dashed line.}
\end{figure}

In Section \ref{agn_sec} we argued that the increase in LIRG fraction
was not dominated by the increasing AGN fraction in our close pairs
sample (Ellison et al. 2011b).  The enhancement in the SFR in pairs is
therefore the obvious culprit responsible for the increasing LIRG
fraction at small separations.  In Figure \ref{sfr} we show the
distribution of SFRs for the pair and control samples, distinguishing
between those that are classified as LIRGs and those with lower \LIR.
In the top and middle panels we compare the LIRG (solid lines) and
non-LIRG (dot-dashed lines) SFRs for control (top) and pairs (middle).
It can be seen that the SFRs of the LIRGs are shifted to higher
values, in both the pairs and control, relative to the SFRs in the
non-LIRGs.  In the lower panel, we compare the SFRs of the pair LIRGs
(red) with the control LIRGs (black).  The SFRs are indistinguishable;
the controls that are LIRGs have similarly high SFRs to those pairs
that are LIRGs.  
Figure \ref{sfr} indicates that whilst the interaction serves to
increase the SFR, and hence the \LIR, similarly high values of SFR
(and \LIR) can be found amongst the control sample.  This result is
in contrast with the ULIRGs, where the high SFRs are almost exclusively
associated with mergers.

Star formation alone does not produce high IR fluxes; it is the presence
of dust to re-process the shorter wavelengths that is the second
key ingredient in the production of a LIRG.  In Figure \ref{ebmv}
we present distributions of \ebmv\ for the LIRG/non-LIRG pairs and
control in the same fashion as for SFRs in Figure \ref{sfr}.  The
\ebmv\ values are determined from the Balmer decrement using an
SMC extinction curve (see Scudder, Ellison \& Mendel 2012 for full details of
our emission line processing).  As was the case for SFRs, we find that
the LIRGs in both the pair and control samples have higher \ebmv\
than the non-LIRGs.  Moreover, the pair LIRGs and control LIRGs
(lower panel) have similar \ebmv\ distributions.  A high extinction
may be due to an intrinsically higher dust content, or it can
be an inclination effect.  We investigated the disk inclinations
derived for our sample from the public SDSS photometric catalogue of
Simard et al. (2011), but found no difference between the LIRG and
non-LIRG distributions.  The high \LIR\ in LIRGs is therefore apparently
not an inclination effect, but relies on intrinisically high SFRs
in the presence of dust.

\subsection{How to make a LIRG}\label{how2_sec}

It is clear from the results presented in this paper, and elsewhere
(e.g. Robaina et al. 2009; Kartaltepe et al. 2010; Hwang et al. 2010,
2011), that mergers contribute to the LIRG population.  In this paper,
we have demonstrated the LIRG-merger connection in two main ways.
First, we have shown that the LIRG fraction increases in our sample of
close pairs as the galaxies' projected separation decreases (Figure
\ref{lirg_frac}).  Second, we have shown in Figure \ref{merger_frac}
that the contribution from mergers increases at IR luminosities above
the LIRG cut-off.  We have also demonstrated that equal mass pairings
of massive, star-forming galaxies are particularly conducive to
producing LIRGs.  However, as demonstrated in Figure
\ref{merger_frac}, mergers are clearly not \textit{necessary} for the
production of LIRGs; any process which enables the galaxy to reach a
high enough SFR (in the presence of dust) may lead to a LIRG.  This
logic implies that there is some SFR threshold required for a galaxy
to appear as a LIRG.  Indeed, as shown in Figure \ref{sfr} there are
very few LIRGs with SFR $<$ 1 M$_{\odot}$/yr.  This is not surprising,
given the strong correlation between \LIR\ and SFR (e.g. Kennicutt
1998; Kennicutt \& Evans 2012).  Forming a LIRG could therefore be as
simple as achieving sufficient SFR (in the presence of dust).

We can approximately\footnote{The finer details of a galaxy's \LIR\
will depend on factors other than just a SFR.  Dust must obviously be
present, and the spatial distribution of stars and dust may also play
a role.} determine the SFR threshold required to produce log (\LIR/\lsun) $>$
11 empirically from our data by performing a linear least
squares fit between the \LIR\ values from the H10 catalog and the SFRs from
the matched SDSS spectroscopic sample.  We find that log (\LIR/\lsun) = 11
corresponds to a SFR $\sim$ 9 M$_{\odot}$/yr.  Alternatively,
from the \LIR-mass relation (that results from correlations between
SFR-\LIR\ and SFR-mass) we find that a log (\LIR/\lsun) = 11 corresponds
to a typical mass log (M$_{\star}$/\msun) $\sim$ 10.9, at $z \sim 0$.
`Normal' star-forming galaxies above log (M$_{\star}$/\msun)  = 10.9 may
therefore readily manifest themselves as LIRGs, \textit{modulo} the
observed scatter in the stellar mass-SFR relation for individual
galaxies.  Lower mass galaxies require a SFR enhancement above the
average for their stellar mass.  This enhancement is simply the
difference between a SFR of 9 M$_{\odot}$/yr and the value predicted
from the stellar mass-SFR relation and is determined to be

\begin{equation}\label{lirg_delta_eqn}
\Delta \rm{SFR} = 8.709 - 0.797 \log M_{\star}.
\end{equation}

The quantity $\Delta$SFR is identical to the SFR offsets that we have
previously calculated for the galaxy pairs in Scudder et al. (2012).
That is, for a given galaxy in a pair, the difference between its SFR
and the median of its mass-matched controls.  For the present work, we
have expanded the calculation of $\Delta$SFRs beyond just the pairs
sample, to include all galaxies with measured masses in the DR7
spectroscopic sample (i.e. the sample presented in Mendel et
al. 2013).  The $\Delta$SFR of a given galaxy is calculated relative
to all galaxies on the star-forming `main sequence' (i.e. the
SFR-stellar mass relation) within a redshift tolerance $\Delta
z$=0.005 and mass tolerance $\Delta$log M$_{\star}$=0.1 dex.  At least 5
matches are required in order to calculate a SFR offset, a limit which
is achieved in 93\% of cases (the typical number of mass and redshift
matches is usually of the order several hundred).  If less than 5
matches are found, the tolerance is grown by a further $\Delta
z$=0.005 in redshift and $\Delta$log M$_{\star}$=0.1 dex in stellar mass
until the required number of matches is achieved.

With SFR offsets in hand, it is possible to directly compare the
$\Delta$SFR of the LIRG pairs with the relation in equation
\ref{lirg_delta_eqn}.  In the top panel of Figure \ref{delta_sfr} we
plot the $\Delta$SFRs of the pairs vs their stellar mass; large
circles indicate galaxies identified as LIRGs (after the flux is
aportioned relative to SFR) and small points show non-LIRG galaxies.
The diagonal dashed line shows the minimum $\Delta$SFR required in
order to produce a LIRG (from equation \ref{lirg_delta_eqn}).  The
lower panel also depicts the $\Delta$SFR versus mass, but for the
control galaxies. Despite its simplicity (and the significant scatter
in the \LIR-SFR relation, and smaller scatter in the stellar mass-SFR
relation) equation \ref{lirg_delta_eqn} is a fairly good
representation of the data, with most of the LIRGs lying above the
diagonal dashed line in both panels of Fig. \ref{delta_sfr}, and
most non-LIRGs lying at smaller values of $\Delta$SFR for their stellar
mass.  Nonetheless, there are some non-LIRG galaxies whose $\Delta$
SFR appears, at face value, to be sufficient to produce a LIRG.
The non-LIRGs in the shaded regions of Figure \ref{delta_sfr}
do have \ebmv\ values that are lower, on average, than the LIRGs,
but there is significant overlap.  Presumably, the location of the
dust relative to the starburst is important for the production of
large IR fluxes.

It can be seen that, in both the pairs and control samples, 
LIRGs are present across the full mass range, but
larger $\Delta$SFRs are required at lower masses, by a factor of 10 or
more, as predicted from equation \ref{lirg_delta_eqn}.  SFR offsets of
a factor of 10 are very rare; Scudder et al. (2012) find only 5\% of
pairs have log $\Delta$ SFR $>$ 1.  Figure \ref{delta_sfr} shows that
SFR offsets of a factor of 10 are completely absent from the
control sample.  The fact that LIRGs are more common at higher mass
(e.g. Melbourne et al. 2008; Hwang et al. 2010), is presumably due, at
least in part, to the lower $\Delta$SFR that is required to put them
above the LIRG threshold (diagonal dashed line in Figure \ref{delta_sfr}).  
Put another way, it is not that mergers between massive galaxies
necessarily produce the largest SFR enhancements (in fact, Figure
\ref{delta_sfr} shows a spread of $\Delta$SFR at fixed mass), but
rather they require relatively small boosts (or no boost at all!)  in
order to achieve a LIRG-sufficient SFR.  Indeed, the fraction of LIRGs
with peculiar morphologies (presumably associated with an interaction)
increases towards lower redshifts (Melbourne et al. 2005).  Secular
processes, such as bars, which can enhance SFRs by similar amounts to
mergers (e.g. Ellison et al. 2011a and references therein) may play a
significant role in producing low redshift LIRGs (e.g. Wang et
al. 2006).

\subsection{ULIRGs}

In this paper, we have focused on LIRGs, largely because they are the
highest \LIR\ galaxies we can study with reasonable statistics in our
sample.  The one ULIRG in our sample is IRAS F08572+3915 (SDSS
objID=588013382191349914).  This is a well known ULIRG and is the only
galaxy pair in our sample that is present in the IRAS Revised Bright
Galaxy Sample (Sanders et al. 2003).  This galaxy pair is also present
in the Great Observatory All-sky LIRG Survey (GOALS, Armus et
al. 2009).  The IR luminosity for this ULIRG (from Hwang et al. 2010)
is log (\LIR/\lsun) = 12.01, slightly lower than reported in the catalog of
Wang \& Rowan-Robinson (2009) log (\LIR/\lsun) = 12.16.  If we aportion the
IR flux according to the SFRs in each of the two galaxies, as we have
done for the LIRGs, the lower \LIR\ in the Hwang et al. (2010) catalog
yields two LIRGs.  However, the slightly higher \LIR\ in the Wang \&
Rowan-Robinson (2009) is sufficient to preserve the ULIRG
classification of the more highly star-forming component in the pair.

\begin{figure}
\centerline{\rotatebox{0}{\resizebox{8cm}{!}
{\includegraphics{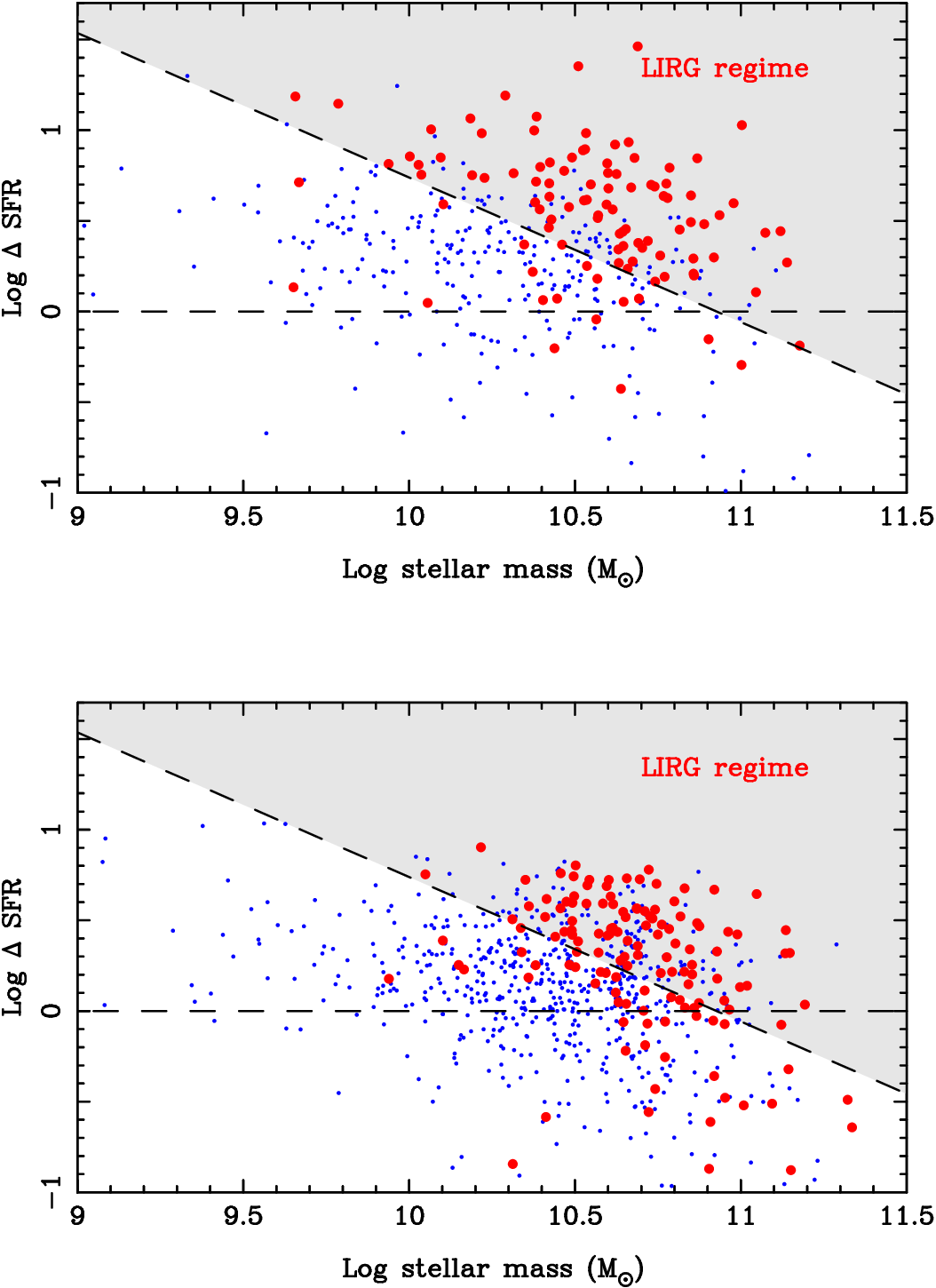}}}}
\caption{\label{delta_sfr} $\Delta$SFRs versus stellar
mass for galaxies classified as LIRGs (large red circles) and non-LIRGs
(small blue points).   The diagonal dashed line shows
the minimum $\Delta$SFR required in order to produce a LIRG (from
equation \ref{lirg_delta_eqn}). The top panel shows galaxies in our
pairs sample; the control galaxies are shown in the lower panel.}
\end{figure}

Despite the lack of ULIRGs in our sample, it is worth considering how
the discussion in the previous section applies to ULIRGs, with log 
(\LIR/\lsun) $\ge$ 12.  To become a ULIRG, the $\Delta$SFR has to be 
higher by a
further order of magnitude than the diagonal dashed line in Figure
\ref{delta_sfr}, which readily explains why ULIRGs are so rare in the
local universe; unlike LIRGs, they are not `naturally occurring' at
any stellar mass.  Even in mergers, the SFR boosts are rarely sufficient
to produce a log (\LIR/\lsun) $>$ 12 in our SDSS sample.
However, at high redshifts, the relationship between mass and SFR
evolves, such that SFRs are higher at fixed stellar mass by a factor
of 30 at $z \sim$ 2 relative to $z \sim 0$ (e.g. Daddi et al. 2007;
Noeske et al. 2007; Wuyts et al. 2011).  This evolution means that a
$\Delta$SFR sufficient to produce a LIRG at $z \sim 0$ (as in our
SDSS sample) is sufficient to produce a ULIRG at $z \sim 2$.  Indeed,
we have (at most) 2 ULIRGs after SFRs are aportioned, $\sim$ 0.5\% of the
sample, compared to 7\% of pairs at $z \sim 1$ (Lin et al. (2007).

There has recently been a shift in our understanding of the role of IR
luminous galaxies and the implications of a classification scheme that
has redshift independent \LIR\ definitions.  It has been noted in
several works that high IR luminosities are the norm at $z \sim 2$
such that ULIRGs at low and high redshift are very different beasts
(e.g. Daddi et al. 2007, 2010; Elbaz et al. 2007, 2011; Sargent et
al. 2012).  Whilst a `normal' M$_{\star} \sim 10^{11}$ \msun\
star-forming galaxy is a LIRG at $z=0$, a similar mass galaxy will be
a ULIRG at $z=2$, and most galaxies with masses above 10$^{10}$ \msun\
will be LIRGs.  Indeed, most ULIRGs are regular main sequence galaxies
by $z \sim 1$ (Sargent et al. 2012).  It is therefore not surprising
that LIRGs dominate the cosmic star formation budget by $z\sim$ 1 (Le
Floc'h et al 2007; Magnelli et al 2009).  Rather than using fixed
limits in \LIR, identifying the galaxies with the most enhanced SFRs
at a given redshift may be best achieved by, as we have done here,
quantifying offsets from the star-forming `main sequence' of galaxies
at the same epoch (see also, e.g., Rodighiero et al. 2011; Nordon et
al. 2012; Sargent et al. 2012; Rovilos et al. 2012).  Nonetheless,
Kartaltepe et al. (2012) have argued that most $z \sim 2$ ULIRGS
\textit{are} still associated with mergers, despite apparently being
in a steady, `maintenance' mode of star formation, rather than
undergoing the intense starbursts asscoiated with the low redshift
ULIRGs (Elbaz et al. 2011).  This could be explained if high redshift
mergers tend to trigger much lower (if any) amounts of star formation
than their low redshift counterparts.  Such a scenario has recently
been suggested by Xu et al. (2012) who do not find an enhancement in
the SFR of mergers at $z > 0.4$ from Spitzer data. Determining the
distribution of SFRs amongst mergers relative to a well matched
control sample is the natural next step in understanding the
contribution of interactions to SFRs at high redshifts.

\section{Summary}

From the SDSS DR7 we have compiled a sample of 9397 galaxies in close
pairs with mass ratios 0.1 $\le$ M$_1$/M$_2$ $\le$ 10, projected
separations \rp $\le 80$ \hkpc\ and velocity differences \dv\ $\le$ 300
\kms.  The pairs sample has been matched to the SDSS-IRAS catalog of
Hwang et al. (2010) in order to investigate the connection between
mergers and high IR luminosities.  Total IR luminosities are measured for
400 of the pair galaxies, of which 172 are LIRGs and 1 is a ULIRG.  For
comparison, a mass-, redshift- and environment-matched control sample
of 4 galaxies per pair galaxy is also compiled.   Our main conclusions
are as follows.

\begin{enumerate}

\item The LIRG fraction of galaxy pairs
increases with decreasing projected separation relative to the control
sample (Figure \ref{lirg_frac}).  Such an increase is
consistent with expectations that interactions are triggering star
formation which, in turn, increase the IR luminosity.  Interestingly,
we find an increased LIRG fraction out to at least 80 \hkpc\
(the limit of our sample), whereas
many previous LIRG studies have focused on morphological
classifications and late stage mergers close to coalescence.

\item Most of the LIRG pairs in our sample are biased to high mass
ratios (Figure \ref{props}), due to the higher SFRs that are
intrinsically found at higher masses.  The highest IR luminosities are
therefore not necessarily found in galaxies with the highest SFR
\textit{enhancements}.  

\item The pairs with the highest \LIR\ are generally
those with the highest total mass, where the \LIR\ reflects the total
SFR in the IRAS beam.  However, we do find that at fixed total stellar
mass ($M_2+M_2$), major mergers (0.33 $< M_1/M_2 <$ 3) have, on
average, slightly higher \LIR\ than minor mergers (Figure \ref{mtot}).  
The most likely
candidates for (merger-induced) LIRGs are therefore close pairs where
both galaxies are star-forming and have approximately equal, high
stellar masses.

\item LIRG pairs are found in relative low density environments
compared with the full pairs sample.  However, the environments of
LIRGs are consistent with those of other star-forming (IRAS-detected)
pairs.

\item The enhanced \LIR\ in the close pairs relative to the control is
seen in both the star forming and composite (star forming+AGN)
galaxies (Figure \ref{lir_hist_AGN}).  However, there is no statistical
difference between the \LIR\ distributions of the pair AGN and control
AGN.  The enhanced \LIR\ in close pairs is therefore likely to be
driven by star formation rather than a higher incidence of AGN at
close separations.

\item The AGN fraction increases with \LIR\ in both pairs and control
galaxies, although there is marginal evidence that the pair AGN
fraction is lower than the control at fixed \LIR\ (Figure \ref{AGN_frac}).

\item The fraction of galaxies classified as close pairs increases
with \LIR\ above log (\LIR/\lsun) $\sim$ 11 (Figure
\ref{merger_frac}).  Close pairs (\rp $\le$ 30 \hkpc, 0.33 $< M_1/M_2 <$ 3)
represent 7\% of the LIRG population.

\item LIRGs have higher SFRs and \ebmv\ than non-LIRGs (Figures
\ref{sfr} and \ref{ebmv}).  This is true for both pair and control
galaxies.  In fact, the SFRs and \ebmv\ distributions are
indistinguishable for LIRGs with/without a close companion.

\item We define a SFR offset, $\Delta$SFR, as the difference
between a galaxy's SFR and the median of other star-forming galaxies
at equivalent mass and redshift.  The $\Delta$SFR required to become a
LIRG decreases as mass increases.  A regular star-forming galaxy with
mass above log (M$_{\star}$/\msun) $\sim$ 11 (at $z \sim 0$) is likely to
have sufficient rates of star formation to be classified as a LIRG
without any additional SFR enhancement.  Lower mass galaxies require
higher boosts; mergers can clearly contribute such boosts.

\end{enumerate}

The results presented here demonstrate the link between galaxy mergers
and the high IR luminosities that define the LIRG population.
However, we have also argued that interactions are just one (and,
possibly, not the dominant) route to producing LIRGs.  

\section*{Acknowledgments} 

SLE and DRP acknowledge the receipt of NSERC Discovery grants which
funded this research.  Ho Seong Hwang generously shared his IRAS-SDSS
matched catalog with us.  We are grateful to Andreea Petric for
wavelength-related wisdom and suggestions that motivated this
work, and to the anonymous referee for several useful suggestions.

Funding for the SDSS and SDSS-II has been provided by the Alfred
P. Sloan Foundation, the Participating Institutions, the National
Science Foundation, the U.S. Department of Energy, the National
Aeronautics and Space Administration, the Japanese Monbukagakusho, the
Max Planck Society, and the Higher Education Funding Council for
England. The SDSS Web Site is http://www.sdss.org/.

The SDSS is managed by the Astrophysical Research Consortium for the
Participating Institutions. The Participating Institutions are the
American Museum of Natural History, Astrophysical Institute Potsdam,
University of Basel, University of Cambridge, Case Western Reserve
University, University of Chicago, Drexel University, Fermilab, the
Institute for Advanced Study, the Japan Participation Group, Johns
Hopkins University, the Joint Institute for Nuclear Astrophysics, the
Kavli Institute for Particle Astrophysics and Cosmology, the Korean
Scientist Group, the Chinese Academy of Sciences (LAMOST), Los Alamos
National Laboratory, the Max-Planck-Institute for Astronomy (MPIA),
the Max-Planck-Institute for Astrophysics (MPA), New Mexico State
University, Ohio State University, University of Pittsburgh,
University of Portsmouth, Princeton University, the United States
Naval Observatory, and the University of Washington.

\end{document}